\documentstyle{mn}
\input psfig

\newif\ifAMStwofonts
\AMStwofontstrue

\def\aaa#1{{A\&A,} {#1}}

\def\apj#1{{ApJ,} {#1}}

\def\mnras#1{{MNRAS,} {#1}}

\ifoldfss
  \ifCUPmtlplainloaded \else
    \NewTextAlphabet{textbfit} {cmbxti10} {}
    \NewTextAlphabet{textbfss} {cmssbx10} {}
    \NewMathAlphabet{mathbfit} {cmbxti10} {} % for math mode
    \NewMathAlphabet{mathbfss} {cmssbx10} {} %  "   "    "
  \fi
  \ifAMStwofonts
    \ifCUPmtlplainloaded \else
      \NewSymbolFont{upmath} {eurm10}
      \NewSymbolFont{AMSa} {msam10}
      \NewMathSymbol{\upi}     {0}{upmath}{19}
      \NewMathSymbol{\umu}     {0}{upmath}{16}
      \NewMathSymbol{\upartial}{0}{upmath}{40}
      \NewMathSymbol{\leqslant}{3}{AMSa}{36}
      \NewMathSymbol{\geqslant}{3}{AMSa}{3E}

      \let\leq=\leqslant 
       
    \fi
  \fi
\fi % End of OFSS

\ifnfssone
  \newmathalphabet{\mathit}
  \addtoversion{normal}{\mathit}{cmr}{m}{it}
  \addtoversion{bold}{\mathit}{cmr}{bx}{it}
  \newmathalphabet{\mathbfit} % math mode version of \textbfit{..}
  \addtoversion{normal}{\mathbfit}{cmr}{bx}{it}
  \addtoversion{bold}{\mathbfit}{cmr}{bx}{it}
  \newmathalphabet{\mathbfss} % math mode version of \textbfss{..}
  \addtoversion{normal}{\mathbfss}{cmss}{bx}{n}
  \addtoversion{bold}{\mathbfss}{cmss}{bx}{n}
  \ifAMStwofonts
    \ifCUPmtlplainloaded \else
      \UseAMStwoboldmath
      \makeatletter
      \new@mathgroup\upmath@group
      \define@mathgroup\mv@normal\upmath@group{eur}{m}{n}
      \define@mathgroup\mv@bold\upmath@group{eur}{b}{n}
      \edef\UPM{\hexnumber\upmath@group}
      \new@mathgroup\amsa@group
      \define@mathgroup\mv@normal\amsa@group{msa}{m}{n}
      \define@mathgroup\mv@bold\amsa@group{msa}{m}{n}
      \edef\AMSa{\hexnumber\amsa@group}
      \makeatother
      \mathchardef\upi="0\UPM19
      \mathchardef\umu="0\UPM16
      \mathchardef\upartial="0\UPM40
      \mathchardef\leqslant="3\AMSa36
      \mathchardef\geqslant="3\AMSa3E

      \let\leq=\leqslant 

    \fi
  \fi
\fi % End of NFSS release 1

\ifnfsstwo
  \DeclareMathAlphabet{\mathbfit}{OT1}{cmr}{bx}{it}
  \SetMathAlphabet\mathbfit{bold}{OT1}{cmr}{bx}{it}
  \DeclareMathAlphabet{\mathbfss}{OT1}{cmss}{bx}{n}
  \SetMathAlphabet\mathbfss{bold}{OT1}{cmss}{bx}{n}
  \ifAMStwofonts
    \ifCUPmtlplainloaded \else
      \DeclareSymbolFont{UPM}{U}{eur}{m}{n}
      \SetSymbolFont{UPM}{bold}{U}{eur}{b}{n}
      \DeclareSymbolFont{AMSa}{U}{msa}{m}{n}
      \DeclareMathSymbol{\upi}{0}{UPM}{"19}
      \DeclareMathSymbol{\umu}{0}{UPM}{"16}
      \DeclareMathSymbol{\upartial}{0}{UPM}{"40}
      \DeclareMathSymbol{\leqslant}{3}{AMSa}{"36}
      \DeclareMathSymbol{\geqslant}{3}{AMSa}{"3E}

      \let\leq=\leqslant 

    \fi
  \fi
\fi % End of NFSS release 2

\ifCUPmtlplainloaded \else
  \ifAMStwofonts \else % If no AMS fonts
    \def\upi{\pi}
    \def\umu{\mu}
    \def\upartial{\partial}
  \fi
\fi

\title[Slim accretion discs]
{Slim accretion discs: a model for ADAF-SLE transitions}
\author[Igumenshchev et al.]
{
	Igor V. Igumenshchev$^{1,2}$\thanks{E-mail: ivi@fy.chalmers.se},
	Marek A. Abramowicz$^{1,3,4}$ and  Igor D. Novikov$^{5,3,6,7}$\\
        $^1$Department of Astronomy \& Astrophysics, G{\"o}teborg
           University and Chalmers University of Technology,
           412 96 G{\"o}teborg, Sweden \\
	$^2$Institute of Astronomy, 48 Pyatnitskaya Street, 
 	   Moscow, 109117, Russia \\
	$^3$Nordita, Blegdamsvej 17, DK-2100 Copenhagen \O, Denmark \\
	$^4$Laboratorio Interdisciplinare SISSA, Trieste, Italy, and ICTP,
 	   Trieste, Italy \\
	$^5$Theoretical Astrophysics Center, Juliane Maries Vej 30,
	   DK-2100 Copenhagen \O, Denmark \\
        $^6$University Observatory, Juliane Maries Vej 30,
           DK-2100 Copenhagen \O, Denmark \\
        $^7$P. N. Lebedev Physical Institute, 84/32 Profsoyuznaya Street,
           Moscow, 117810, Russia \\
}

\date{Accepted 1997 September 00.
      Received 1997 September 00;
      in original form 1997 September 00}

\pagerange{\pageref{firstpage}--\pageref{lastpage}}
\pubyear{1997}

\begin{document}

\maketitle

\label{firstpage}

\begin{abstract}

We numerically construct slim, global, vertically integrated models of
optically thin, transonic accretion discs around black holes, assuming
a regularity condition at the sonic radius and boundary conditions
at the outer radius of the disc and near the black hole.
%boundary conditions at the sonic radius, at the outer radius of the disc
%and near the black hole.
In agreement with several previous studies, we
find two branches of shock-free solutions, in which the cooling is
dominated either by advection, or by local radiation. We also confirm
that the part of the accretion flow where advection dominates is in some
circumstances limited in size: it does not extend beyond a certain
outer limiting radius. New results found in our paper concern the
location of the limiting radius and properties of the flow near to it. In
particular, we find that beyond the limiting radius, the advective
dominated
solutions match on to Shapiro, Lightman \& Eardley (SLE) discs
through a smooth transition region. Therefore, the full
global solutions are shock-free and unlimited in size. There is no need
for postulating an extra physical effect (e.g. evaporation) for
triggering the ADAF-SLE transition. It occurs due to standard accretion
processes described by the classic slim disc equations.

\end{abstract}


\section{Introduction}

%        \footnote[0]{$^\star$E-mail: illarion@dpc.asc.rssi.ru (AFI);
%         ivi@fy.chalmers.se (IVI)}

Most of our knowledge about accretion disc structure comes from a very
successful, but simplified, approach introduced by Shakura (1972),
Pringle \& Rees (1972), Shakura \& Sunyaev (1973) and others. In this
approach, only the radial structure of the disc is studied in a
detailed way, and all of the disc properties are taken to be
independent of time and 
azimuthal angle and are averaged in the vertical direction. Let $H = H(r)$
be the vertical
extension of the disc at a particular radial location $r$. In the
standard Shakura-Sunyaev `thin' model, terms of the order 
$(H/r)^2$ and higher are neglected.  This allows the problem
to be reduced to a set of linear algebraic equations which it is easy to
solve: the standard model gives explicitly all of physical characteristics 
of the
disc.

However, the $(H/r)^2$ terms, which are neglected in the standard thin
disc
model, describe effects that in many astrophysical applications are
very relevant: the radial pressure gradient, non-Keplerian angular
momentum
distribution, and cooling by radial advection of heat. These terms are
all retained in the `slim' accretion discs models introduced by
Abramowicz et al. (1988, hereafter ACLS88) and based on a set of
non-linear radial differential equations derived by Paczy{\'n}ski \&
Bisnovatyi-Kogan (1981). Mathematically, integrating the slim disc
equations leads to a non-standard eigenvalue problem which is
surprisingly difficult to solve.  The original ACLS88 slim disc models,
and slim models studied later by many other authors, have been
constructed with a help of a very accurate pseudo-Newtonian model of
the black hole gravitational field (Paczy{\'n}ski \& Wiita 1980). Lasota
(1994) derived the full set of slim accretion disc equations in
Kerr geometry, i.e. in the gravitational field of a rotating
black hole. Abramowicz et al. (1996, hereafter ACGL96), and
Abramowicz, Lanza \& Percival (1997) later made some improvements and
corrections to Lasota's equations. Recently, Beloborodov, Abramowicz \&
Novikov (1997) rederived the  relativistic slim accretion disc
equations again, taking into account the relativistic effect of the
heat inertia that is important for discs around rapidly rotating black
holes, but was neglected by previous investigators.

Initially, all of the slim models were constructed under the
assumption of the disc being very optically thick but a few years ago it
was realized by Narayan \& Yi (1994), Abramowicz et al. (1995),
and Narayan \& Yi (1995) that a new class of very hot, optically thin,
advective discs exists. The new class are called ADAFs and are very
promising for explaining properties of X-ray transients, low luminosity
active galactic nuclei, the Galactic Centre, and other high energy
objects.
For this reason they are now being studied intensively by many
researchers (see
Narayan 1997, for a review and references).

The first ADAF models were local (in the sense that no boundary
conditions were applied), and were constructed with many simplifying
assumptions. Only very recently have several authors constructed more
accurate, slim global models. Chen, Abramowicz \& Lasota (1997),
Narayan, Kato \& Honma (1997), and Nakamura et al. (1997)
constructed global transonic models for ADAFs using the
Paczy{\'n}ski-Witta model for the black hole gravity. ACGL96 and
Jaroszy{\'n}ski \& Kurpiewski (1997) constructed global transonic ADAF
models in Kerr geometry, using Lasota's equations.

In this paper we construct some new black hole models of slim discs,
using Lasota's equations in the version given by ACGL96 and Abramowicz
et al. (1997a), where the influence of the heat inertia
(discussed in Beloborodov et al. 1997) was neglected. Equations,
boundary conditions and the numerical procedure are presented in
Section~2.
In Section~3 we give the numerical results for adiabatic and
non-adiabatic models. Section~4 contains the final discussion and
conclusions. The new type of global transonic models found in our paper
consist of an ADAF in the inner part and an SLE (Shapiro, Lightman \&
Eardley 1976) type of flow in the outer part. The SLE discs are
known to be violently unstable (Piran 1978), and therefore our
solutions with the ADAF-SLE transitions are not astrophysically
realistic models of stationary accretion flows. However, they provide a
non-trivial example of the ADAF-(thin disc) transition and for this
reason they could be of interest.

\section[]{METHOD OF SOLUTION}

\setcounter{equation}{0}

\subsection{Equations}

The equations for Keplerian discs in the Kerr metric were derived by
Novikov \& Thorne (1973).
We follow the notation and equations in ACGL96 for non-Keplerian
slim discs, except where explicitly indicated.

\noindent
The mass conservation equation reads,
%%%%%%%%%%%%%%%%%%%%%%%%%%%%%%%%%%%%%%%%%%%%%%%%%%%%%%%%%%%%%%%%%%%%%%%%%%%%%%%
\begin{equation}
   \dot{M}=-2\pi\Delta^{1/2}\Sigma{V\over\sqrt{1-V^2}}, 
\end{equation}
%%%%%%%%%%%%%%%%%%%%%%%%%%%%%%%%%%%%%%%%%%%%%%%%%%%%%%%%%%%%%%%%%%%%%%%%%%%%%%%
where $\Sigma=2H\rho$ is the surface density, $H$ is the half-thickness
of the disc, $\rho$ is the rest mass density at the equatorial
plane and $V$ is the radial velocity, defined later by the equation
(2.16).

\noindent
The equation of radial momentum conservation takes the form,
%%%%%%%%%%%%%%%%%%%%%%%%%%%%%%%%%%%%%%%%%%%%%%%%%%%%%%%%%%%%%%%%%%%%%%%%%%%%%%%
\begin{equation}
  {{V\over\sqrt{1-V^2}}}{dV\over dr}={{\cal A}\over r}-
   {1\over\rho}{dp\over dr},
\end{equation}
%%%%%%%%%%%%%%%%%%%%%%%%%%%%%%%%%%%%%%%%%%%%%%%%%%%%%%%%%%%%%%%%%%%%%%%%%%%%%%%
where the pressure $p$ is at the equatorial plane, and
%%%%%%%%%%%%%%%%%%%%%%%%%%%%%%%%%%%%%%%%%%%%%%%%%%%%%%%%%%%%%%%%%%%%%%%%%%%%%%%
\begin{equation}
   {\cal A}=-{MA\over r^3\Delta\Omega_K^+\Omega_K^-}
   {(\Omega-\Omega_K^+)(\Omega-\Omega_K^-)\over
   1-\tilde{\Omega}^2\tilde{R}^2}.
\end{equation}
%%%%%%%%%%%%%%%%%%%%%%%%%%%%%%%%%%%%%%%%%%%%%%%%%%%%%%%%%%%%%%%%%%%%%%%%%%%%%%%
The definitions of the symbols introduced above are given later in this
Section.
The last term in equation (2.2) differs from the one used by
ACGL96.

\noindent
The equation of angular momentum conservation,
%%%%%%%%%%%%%%%%%%%%%%%%%%%%%%%%%%%%%%%%%%%%%%%%%%%%%%%%%%%%%%%%%%%%%%%%%%%%%%%
\begin{equation}
   {\dot{M}\over 2\pi r}{d{\cal L}\over dr}+{1\over r}{d\over dr}\left(
   \Sigma\nu A^{3/2}{\Delta^{1/2}\gamma^3\over r^4}{d\Omega\over dr}
   \right)=0,
\end{equation}
%%%%%%%%%%%%%%%%%%%%%%%%%%%%%%%%%%%%%%%%%%%%%%%%%%%%%%%%%%%%%%%%%%%%%%%%%%%%%%%
where
%%%%%%%%%%%%%%%%%%%%%%%%%%%%%%%%%%%%%%%%%%%%%%%%%%%%%%%%%%%%%%%%%%%%%%%%%%%%%%%
\begin{equation}
  {\cal L}=-u_\varphi=\gamma\left({A^{3/2}\over r^3\Delta^{1/2}}\right)
   \tilde{\Omega}
\end{equation}
%%%%%%%%%%%%%%%%%%%%%%%%%%%%%%%%%%%%%%%%%%%%%%%%%%%%%%%%%%%%%%%%%%%%%%%%%%%%%%%
is the specific angular momentum.

\noindent
The equation of energy conservation in the one temperature ($T_e = T_i
=T$) approximation,
%%%%%%%%%%%%%%%%%%%%%%%%%%%%%%%%%%%%%%%%%%%%%%%%%%%%%%%%%%%%%%%%%%%%%%%%%%%%%%%
\begin{equation}
   {\dot{M}\over 2\pi r^2}{p\over\rho}\left({1\over \gamma_g-1} {d\ln
   T\over d\ln r}-{d\ln\rho\over d\ln r}\right)=F^+-F^-,
\end{equation}
%%%%%%%%%%%%%%%%%%%%%%%%%%%%%%%%%%%%%%%%%%%%%%%%%%%%%%%%%%%%%%%%%%%%%%%%%%%%%%%
where $\gamma_g$ is the adiabatic index of the gas,
%%%%%%%%%%%%%%%%%%%%%%%%%%%%%%%%%%%%%%%%%%%%%%%%%%%%%%%%%%%%%%%%%%%%%%%%%%%%%%%
\begin{equation}
   F^+=\nu\Sigma{A^2\over r^6}\gamma^4\left({d\Omega\over dr} \right)^2
\end{equation}
%%%%%%%%%%%%%%%%%%%%%%%%%%%%%%%%%%%%%%%%%%%%%%%%%%%%%%%%%%%%%%%%%%%%%%%%%%%%%%%
is the surface heat generation rate, and $F^-$ is the radiative cooling
flux. For the case of proton-electron bremsstrahlung cooling, the
emissivity per unit volume takes the form
%%%%%%%%%%%%%%%%%%%%%%%%%%%%%%%%%%%%%%%%%%%%%%%%%%%%%%%%%%%%%%%%%%%%%%%%%%%%%%%
\begin{equation}
   {F^-\over 2H}= 5.6\times10^{20}\rho^2 T^{1/2}\ ergs\
    cm^{-3}\,s^{-1}. 
\end{equation}
%%%%%%%%%%%%%%%%%%%%%%%%%%%%%%%%%%%%%%%%%%%%%%%%%%%%%%%%%%%%%%%%%%%%%%%%%%%%%%%
The equation (2.6) is written for a gas-pressure dominated medium
with the equation of state
%%%%%%%%%%%%%%%%%%%%%%%%%%%%%%%%%%%%%%%%%%%%%%%%%%%%%%%%%%%%%%%%%%%%%%%%%%%%%%%
\begin{equation}
   p={{\cal R}\over\mu}\rho T ,
\end{equation}
%%%%%%%%%%%%%%%%%%%%%%%%%%%%%%%%%%%%%%%%%%%%%%%%%%%%%%%%%%%%%%%%%%%%%%%%%%%%%%%
where $\cal R$ is the gas constant, $\mu$ is the mean molecular weight
of gas ($\mu=1/2$ for a hydrogen plasma).

\noindent
We take the equation of vertical balance in the form given by
Abramowicz, Lanza \& Percival (1997),
%%%%%%%%%%%%%%%%%%%%%%%%%%%%%%%%%%%%%%%%%%%%%%%%%%%%%%%%%%%%%%%%%%%%%%%%%%%%%%%
\begin{equation}
   {p\over\rho}={1\over 2}{u_\varphi^2-u_t^2a^2+a^2\over r^2}
    \left({H\over r}\right)^2, 
\end{equation}
%%%%%%%%%%%%%%%%%%%%%%%%%%%%%%%%%%%%%%%%%%%%%%%%%%%%%%%%%%%%%%%%%%%%%%%%%%%%%%%
where
%%%%%%%%%%%%%%%%%%%%%%%%%%%%%%%%%%%%%%%%%%%%%%%%%%%%%%%%%%%%%%%%%%%%%%%%%%%%%%%
\begin{equation}
   u_t^2={\gamma^2A\over r^2\Delta}\left({r^2\Delta\over A}+ {A\over
   r^2}\omega\tilde{\Omega}\right)^2 \;{\rm and}\,\;\;
   u_\varphi^2={\gamma^2A\over r^2\Delta}{A^2\over r^4} \tilde{\Omega}^2.
\end{equation}
%%%%%%%%%%%%%%%%%%%%%%%%%%%%%%%%%%%%%%%%%%%%%%%%%%%%%%%%%%%%%%%%%%%%%%%%%%%%%%%

In equations (2.1) -- (2.11), we have used the following definitions,
%%%%%%%%%%%%%%%%%%%%%%%%%%%%%%%%%%%%%%%%%%%%%%%%%%%%%%%%%%%%%%%%%%%%%%%%%%%%%%%
\begin{equation}
\begin{array}{ll}
  \!\!\!\Delta  =  r^2-2Mr+a^2,\quad & 
  \omega  =  {\displaystyle {2Mar\over A}}, \\
  \!\!\!\tilde{R}^2  = {\displaystyle {A^2\over r^4\Delta}},\quad &
  A  =  r^4+r^2a^2+2Mra^2, \\
\end{array}
\end{equation}
%%%%%%%%%%%%%%%%%%%%%%%%%%%%%%%%%%%%%%%%%%%%%%%%%%%%%%%%%%%%%%%%%%%%%%%%%%%%%%%
where $M$ is the mass and $a$ is the total specific angular momentum of
the Kerr black hole.
%%%%%%%%%%%%%%%%%%%%%%%%%%%%%%%%%%%%%%%%%%%%%%%%%%%%%%%%%%%%%%%%%%%%%%%%%%%%%%%
\begin{equation}
  \Omega={u^\varphi\over u^t}\;\;\;\;{\rm and}\;\;\;\;
  \tilde{\Omega}=\Omega-\omega 
\end{equation}
%%%%%%%%%%%%%%%%%%%%%%%%%%%%%%%%%%%%%%%%%%%%%%%%%%%%%%%%%%%%%%%%%%%%%%%%%%%%%%%
are the angular velocities with respect to a stationary observer and to a
local inertial observer, $u^\varphi$ and $u^t$ are 
components of the four-velocity $u^i$ of matter.  The angular
frequencies of the corotating ($+$) and counterrotating ($-$) Keplerian
orbits are
%%%%%%%%%%%%%%%%%%%%%%%%%%%%%%%%%%%%%%%%%%%%%%%%%%%%%%%%%%%%%%%%%%%%%%%%%%%%%%%
\begin{equation}
  \Omega_K^+={M^{1/2}\over r^{3/2}+ aM^{1/2}},\;\;\;\;\;\;\;\;
  \Omega_K^-=-{M^{1/2}\over r^{3/2}- aM^{1/2}}. 
\end{equation}
%%%%%%%%%%%%%%%%%%%%%%%%%%%%%%%%%%%%%%%%%%%%%%%%%%%%%%%%%%%%%%%%%%%%%%%%%%%%%%%
The relation between the Boyer-Lindquist and physical velocity
components in the azimuthal direction is
%%%%%%%%%%%%%%%%%%%%%%%%%%%%%%%%%%%%%%%%%%%%%%%%%%%%%%%%%%%%%%%%%%%%%%%%%%%%%%%
\begin{equation}
   v^{(\varphi)}=\tilde{R}\tilde{\Omega}. 
\end{equation}
%%%%%%%%%%%%%%%%%%%%%%%%%%%%%%%%%%%%%%%%%%%%%%%%%%%%%%%%%%%%%%%%%%%%%%%%%%%%%%%
The radial velocity $V$ is defined by the formula
%%%%%%%%%%%%%%%%%%%%%%%%%%%%%%%%%%%%%%%%%%%%%%%%%%%%%%%%%%%%%%%%%%%%%%%%%%%%%%%
\begin{equation}
  {V\over\sqrt{1-V^2}}=u^r g_{rr}^{1/2},
\end{equation}
%%%%%%%%%%%%%%%%%%%%%%%%%%%%%%%%%%%%%%%%%%%%%%%%%%%%%%%%%%%%%%%%%%%%%%%%%%%%%%%
where $g_{rr}$ is the metric tensor component. The
velocity $V$
is the radial velocity of the fluid as measured by an observer at fixed
$r$ who corotates with the fluid.  The Lorentz gamma factor can be
written as
%%%%%%%%%%%%%%%%%%%%%%%%%%%%%%%%%%%%%%%%%%%%%%%%%%%%%%%%%%%%%%%%%%%%%%%%%%%%%%%
\begin{equation}
  \gamma^2=\left({1\over 1-(v^{(\varphi)})^2}\right) \left({1\over
1-V^2}\right). 
\end{equation}
%%%%%%%%%%%%%%%%%%%%%%%%%%%%%%%%%%%%%%%%%%%%%%%%%%%%%%%%%%%%%%%%%%%%%%%%%%%%%%%
We use the standard assumption for the viscosity coefficient,
%%%%%%%%%%%%%%%%%%%%%%%%%%%%%%%%%%%%%%%%%%%%%%%%%%%%%%%%%%%%%%%%%%%%%%%%%%%%%%%
\begin{equation}
   \nu=\alpha c_s H, 
\end{equation}
%%%%%%%%%%%%%%%%%%%%%%%%%%%%%%%%%%%%%%%%%%%%%%%%%%%%%%%%%%%%%%%%%%%%%%%%%%%%%%%
where $\alpha$ is a parameter and $c_s=(p/\rho)^{1/2}$ is the
isothermal sound speed.

\subsection{Conditions at the sonic radius}

The differential equations (2.2) and (2.6) can be rewritten in a form
where just two unknown functions $V(r)$, and $T(r)$ are present,
%%%%%%%%%%%%%%%%%%%%%%%%%%%%%%%%%%%%%%%%%%%%%%%%%%%%%%%%%%%%%%%%%%%%%%%%%%%%%%%
\begin{equation}
   {d\ln V\over d\ln r}=(1-V^2){N\over D}, 
\end{equation}
%%%%%%%%%%%%%%%%%%%%%%%%%%%%%%%%%%%%%%%%%%%%%%%%%%%%%%%%%%%%%%%%%%%%%%%%%%%%%%%
\[
   {d\ln T\over d\ln r}=-{\gamma_g-1\over\gamma_g+1}\left\{2\left(1-
   \eta V^2\right){N\over D} + \right. \]
\begin{equation}
   \left.\qquad\qquad\qquad{4\pi r^2\over\dot{M}c_s^2}(F^+-F^-)+{\cal
   B}\right\},
\end{equation}
%%%%%%%%%%%%%%%%%%%%%%%%%%%%%%%%%%%%%%%%%%%%%%%%%%%%%%%%%%%%%%%%%%%%%%%%%%%%%%%
where
%%%%%%%%%%%%%%%%%%%%%%%%%%%%%%%%%%%%%%%%%%%%%%%%%%%%%%%%%%%%%%%%%%%%%%%%%%%%%%%
\begin{equation}
   N={\cal A}+{2\pi r^2\over\dot{M}}{\gamma_g-1\over\gamma_g+1}
   (F^+-F^-)+{\gamma_g\over\gamma_g+1}c_s^2{\cal B}, 
\end{equation}
%%%%%%%%%%%%%%%%%%%%%%%%%%%%%%%%%%%%%%%%%%%%%%%%%%%%%%%%%%%%%%%%%%%%%%%%%%%%%%%
\begin{equation}
   D=V^2\left(1+{2\gamma_g\over\gamma_g+1}\eta c_s^2\right)-
   {2\gamma_g\over\gamma_g+1}c_s^2, 
\end{equation}
%%%%%%%%%%%%%%%%%%%%%%%%%%%%%%%%%%%%%%%%%%%%%%%%%%%%%%%%%%%%%%%%%%%%%%%%%%%%%%%
$\cal A$ is defined by equation (2.3) and
%%%%%%%%%%%%%%%%%%%%%%%%%%%%%%%%%%%%%%%%%%%%%%%%%%%%%%%%%%%%%%%%%%%%%%%%%%%%%%%
\[
   {\cal B}= 4 + 2{r(r-M)\over\Delta} -
    \eta\left({2(v^{(\varphi)})^2\over 1-(v^{(\varphi)})^2}{d\ln
    v^{(\varphi)}\over d\ln r}+{d\ln\Psi\over d\ln r}\right),
\]
\[
\Psi=(1-a^2\omega^2){A\over r^2}(v^{(\varphi)})^2-
2a^2\Delta^{1/2}\omega\,v^{(\varphi)}-a^2{r^2\Delta\over A},
\]
\[
\eta=\left(1+{a^2\over\Psi\gamma^2}\right)^{-1}.
\]
%%%%%%%%%%%%%%%%%%%%%%%%%%%%%%%%%%%%%%%%%%%%%%%%%%%%%%%%%%%%%%%%%%%%%%%%%%%%%%%
The regularity condition at the critical point $r_s$ of equations (2.19) and
(2.20) requires that
%%%%%%%%%%%%%%%%%%%%%%%%%%%%%%%%%%%%%%%%%%%%%%%%%%%%%%%%%%%%%%%%%%%%%%%%%%%%%%%
\begin{equation}
  \left. N \right|_{r_s}=0, \;\;\;\;\; \left. D \right|_{r_s}=0,
\end{equation}
%%%%%%%%%%%%%%%%%%%%%%%%%%%%%%%%%%%%%%%%%%%%%%%%%%%%%%%%%%%%%%%%%%%%%%%%%%%%%%%
where $N$ and $D$ are given by the expressions (2.21) and (2.22). In
the nonrelativistic limit, the condition $D=0$ reduces to $
V^2=2\gamma_g c_s^2/(\gamma_g+1)$.  We note that in our version of the
vertically integrated equations, the velocity at the critical point does
not coincide with the adiabatic sound speed, as it does in the case of
the spherical accretion (Bondi 1954).  Nevertheless, following 
tradition, we use the name the `sonic radius' for the critical point
$r_s$, and define the Mach number as
%%%%%%%%%%%%%%%%%%%%%%%%%%%%%%%%%%%%%%%%%%%%%%%%%%%%%%%%%%%%%%%%%%%%%%%%%%%%%%%
\begin{equation}
  {\cal M}=\left({\gamma_g+1\over 2\gamma_g}+\eta c_s^2\right)^{1/2}
  {V\over c_s}. 
\end{equation}
%%%%%%%%%%%%%%%%%%%%%%%%%%%%%%%%%%%%%%%%%%%%%%%%%%%%%%%%%%%%%%%%%%%%%%%%%%%%%%%

\subsection{Boundary conditions}

The set of equations to be solved consists of two first order
differential equations (2.2) and (2.6), and one second order equation
(2.4). To find a solution with given $\dot{M}$, $\alpha$ and $M$, four
integration constants (free parameters) are required, but the regularity
condition (2.23) at the sonic radius reduces the number of free
parameters in the problem to three. The choice of the three free
parameters should be determined by fixing the boundary conditions. This
may be done in many ways, but only certain particular choices will be
consistent with stable numerical algorithms.

The standard approach is to fix boundary conditions at a large
distance
from the central accreting black hole (e.g. ACLS88, Chen et al. 1997) but
a modification of this was
used by Narayan et al. (1997) who specified
$\Omega=\Omega_K$ and $c_s=10^{-3}\Omega r$ at the outer boundary and
the no-torque condition $d\Omega/dr=0$ at the inner boundary located
between the black hole horizon and the sonic radius. It is well known
that when the standard (or modified standard) boundary conditions
are used, the mathematical problem is an eigenvalue
problem that is numerically very time consuming and difficult to solve.

To avoid this difficulty, Chakrabarti and his collaborators (see
Chakrabarti 1996 for references) introduced a very clever mathematical
trick. They assumed a different set of boundary conditions, which are
applied not only at the outer boundary, but also at the sonic radius and
near to the black hole horizon. In this way, the most difficult part of
the problem~-- finding the eigenvalue~-- is trivially solved: the
eigenvalue may be assumed to have any particular value (in practice
from some continuous range) and the outer boundary conditions
automatically follow from it. In a way, this is putting the cart
before the horse: one tries to match a problem to an already known
solution. While this is obviously a perfectly acceptable mathematical
procedure, there is no guarantee that solutions constructed in this way
will always be astrophysically acceptable. One should worry about the
appearance of 
unphysical discontinuities which prevent solutions extending smoothly
out to
very large radii. Troubled solutions such as these may correspond to {\it
incorrect} choices of eigenvalues in the sense that no astrophysically
acceptable outer
boundary conditions are consistent with them.

In order to see how qualitative differences in the treatment of the
boundary conditions influence numerical models of global transonic
solutions describing slim accretion discs, we have adopted in this
paper a procedure that is very similar to the one used by Chakrabarti
and collaborators. Specifically, we take the position of the sonic radius
$r_s$ as one of the free parameters. The second free parameter is fixed
by the condition that the angular momentum tends asymptotically to a
constant value near to the black hole horizon $r_h$.  We assume that at
the inner boundary $r_{in}$ of the disc, the radial change of angular
momentum is already negligible,
%%%%%%%%%%%%%%%%%%%%%%%%%%%%%%%%%%%%%%%%%%%%%%%%%%%%%%%%%%%%%%%%%%%%%%%%%%%%%%%
\begin{equation}
   \left.{d{\cal L}\over dr}\right|_{r_{in}}=0.
\end{equation}
%%%%%%%%%%%%%%%%%%%%%%%%%%%%%%%%%%%%%%%%%%%%%%%%%%%%%%%%%%%%%%%%%%%%%%%%%%%%%%%
Strictly speaking, in (2.25) one should set $r_{in}=r_h$. In 
practice, however, in numerical calculations one may choose the location
of the inner boundary to be
anywhere between the horizon and the sonic point, 
and the results change only slightly.
We use $r_{in}=(r_h+r_s)/2$ in our numerical calculations.

\noindent
The third parameter is specified at the outer boundary $r_{out}$, by
fixing the angular velocity,
%%%%%%%%%%%%%%%%%%%%%%%%%%%%%%%%%%%%%%%%%%%%%%%%%%%%%%%%%%%%%%%%%%%%%%%%%%%%%%%
\begin{equation}
   \Omega(r_{out})=\Omega_{out},
\end{equation}
%%%%%%%%%%%%%%%%%%%%%%%%%%%%%%%%%%%%%%%%%%%%%%%%%%%%%%%%%%%%%%%%%%%%%%%%%%%%%%%
as a fraction of the Keplerian angular velocity $\Omega_K(r_{out})$.

\subsection{Numerical procedure}

The set of nonlinear differential equations (2.4), (2.19) and (2.20)
are solved numerically on a fixed numerical grid by means of a standard
relaxation technique. This determines the solution by starting from an 
initial estimate which is then improved iteratively.
The original differential equations are
replaced by approximate finite-difference equations.  We used two
overlapping grids, one of which, $\{ r_i\}$, $i=1,\cdots,N$, is
used for the second-order equation (2.4) while the other, $\{
r_{i+1/2}\}$, $i=1,\cdots,N-1$, is used for equations
(2.19) and (2.20).  Variables $\Omega_i$, $i=1,\cdots,N$, and
$V_{i+1/2}$, $T_{i+1/2}$, $i=1,\cdots,N-1$, are defined on grids $\{
r_i\}$ and $\{ r_{i+1/2}\}$, respectively.  It is assumed that the
position of the sonic point $r_s$ coincides with one of the points
$r_{i+1/2}$ at $i=N_s$.

The problem involves $3N-6$ coupled finite-difference (algebraical)
equations which depend on $3N-2$ variables.  This set of equations can
be uniquely solved if we exclude two variables $\Omega_1$ and
$\Omega_N$ using the conditions (2.25) and (2.26), respectively, and
another two variables $V_{N_s+1/2}$ and $T_{N_s+1/2}$ using the
conditions (2.23) at the sonic radius.  Then, the system of $3N-6$
algebraical equations
%%%%%%%%%%%%%%%%%%%%%%%%%%%%%%%%%%%%%%%%%%%%%%%%%%%%%%%%%%%%%%%%%%%%%%%%%%%%%%%
\begin{equation}
   F_k(x_1,x_2,\cdots,x_{3N-6})=0 \;\;\;\;\;\; k=1,2,\cdots,3N-6,
\end{equation}
%%%%%%%%%%%%%%%%%%%%%%%%%%%%%%%%%%%%%%%%%%%%%%%%%%%%%%%%%%%%%%%%%%%%%%%%%%%%%%%
where
%%%%%%%%%%%%%%%%%%%%%%%%%%%%%%%%%%%%%%%%%%%%%%%%%%%%%%%%%%%%%%%%%%%%%%%%%%%%%%%
\[ 
\begin{array}{ll}
\!\! x_k= & \left(\Omega_2,\cdots,\Omega_{N-1},V_{3/2},\cdots,V_{N_s-1/2},
       \right.\\
       & \;\;V_{N_s+3/2},\cdots,V_{N-1/2},T_{3/2},\cdots,T_{N_s-1/2},\\
       & \left.\;\,T_{N_s+3/2},\cdots,T_{N-1/2}\right), 
\end{array}
\]
%%%%%%%%%%%%%%%%%%%%%%%%%%%%%%%%%%%%%%%%%%%%%%%%%%%%%%%%%%%%%%%%%%%%%%%%%%%%%%%
is solved by means of a multidimensional Newton iteration method (see,
for example, Press et al. 1992). In matrix notation, the
procedure of solution for equations ${\bf F}({\bf x})=0$ can be
represented as,
%%%%%%%%%%%%%%%%%%%%%%%%%%%%%%%%%%%%%%%%%%%%%%%%%%%%%%%%%%%%%%%%%%%%%%%%%%%%%%%
\begin{equation}
  {\bf x}_{new}={\bf x}_{old}+\varepsilon\cdot\delta{\bf x},
\;\;\;\;\;\;\;\; 0<\varepsilon\leq 1, 
\end{equation}
%%%%%%%%%%%%%%%%%%%%%%%%%%%%%%%%%%%%%%%%%%%%%%%%%%%%%%%%%%%%%%%%%%%%%%%%%%%%%%%
where
%%%%%%%%%%%%%%%%%%%%%%%%%%%%%%%%%%%%%%%%%%%%%%%%%%%%%%%%%%%%%%%%%%%%%%%%%%%%%%%
\begin{equation}
   \delta{\bf x}=-{\bf J}\cdot{\bf F}. 
\end{equation}
%%%%%%%%%%%%%%%%%%%%%%%%%%%%%%%%%%%%%%%%%%%%%%%%%%%%%%%%%%%%%%%%%%%%%%%%%%%%%%%
Here ${\bf J}$ is the Jacobian matrix, 
$J_{ij}\equiv \partial F_i/ \partial x_j$.
We take a value of the parameter $\varepsilon$ less than $1$ to obtain
convergence if the initial guess is not sufficiently close to the
solution but then, when {\bf x} becomes close to the solution, we use the
full
Newton step, $\varepsilon=1$, to obtain the fastest (quadratic)
convergence.  Finally, we check the degree to which both functions and
variables have converged.

\section[]{RESULTS}

\setcounter{equation}{0}

We have constructed two families of models: a family of {\it
adiabatic models} in which radiative cooling was neglected and a 
family of
{\it bremsstrahlung models} with bremsstrahlung cooling. In
both cases we assumed a Schwarzschild black hole, with $a=0$, and
took the disc matter to consist of hydrogen plasma with $\mu=1/2$.  We
used two values for the adiabatic index in the models, $\gamma_g =
5/3,~4/3$, the first corresponding to gas pressure domination
while the second could represent the case where isotropically
tangled magnetic field provides a major contribution to the total
pressure. It is often argued that in realistic astrophysical accretion
flows, the gas and magnetic pressures are roughly the same. Thus, one
may assume that the effective adiabatic index has a value between $4/3$
and $5/3$.

\subsection{Adiabatic models}

We take $F^-=0$ in equation (2.6) in order to study the pure effect of
advection dominated cooling. For a given value of $\gamma_g$, there are
two main parameters of models: $\alpha$ and $r_s$. Solutions depend only
very weakly on other parameters (see Section 2.3). In particular, the
problem is independent of the mass $M$ after a radial re-scaling with
$r_g=2GM/c^2$. 

We now describe models with adiabatic index $\gamma_g=5/3$. Two types of
numerical solution have been found in this case. The first type, which are
limited in size, can be constructed if $r_s>(r_s)_{crit}$, where
$(r_s)_{crit}$ depends on $\alpha$. They cannot extend in the radial
direction beyond a radius $r_{out}^{max}$ at which the thickness $H$ goes
to zero; it is impossible to construct models with $r_{out}$ larger than
this.  The value of $r_{out}^{max}$ is a decreasing function of $r_s$. 
Our numerical procedure can find only an approximate value for
$r_{out}^{max}$, because $dH/dr$ becomes singular ($dH/dr\rightarrow
-\infty$) when $H\rightarrow 0$, but one can set the outer boundary
condition at $r_{out} < r_{out}^{max}$ and obtain solutions which are just
part of the solution with $r_{out}=r_{out}^{max}$.  All of these models
have the ratio $H/r$ less than one everywhere, and so the condition for
the slim disc approach to be self-consistent ($H/r\la 1$) is satisfied. 
Models with $r_s= (r_s)_{crit}$ have the maximal value of $H/r$
approximately equal to one.  Models of this type can experience a
transition at radius $\approx r_{out}^{max}$ to a thin disc of the type
allowed by the energy equation. In Section~3.2, an example of the
transition of an ADAF to an SLE-type disc is given. 

The second type of solution exists when $r_s<(r_s)_{crit}$. Models of this
type are not size-limited in the radial direction; indeed, their
properties depend strongly on the value of $r_{out}$. In particular, the
maximum value of the ratio $H/r$ is an increasing function of $r_{out}$. 
Typically, these models have $H/r\gg 1$ for the whole range of radii,
$r_{in}<r<r_{out}$ (models with $H/r<1$ everywhere exist only for a small
value of $r_{out}$). This type of solution is not self-consistent because
it does not satisfy the slim disc condition for large $r$ and so we will
not discuss it further. In the following, we will only discuss models of
the first type, which have limited radial size. 

%   ----------- Table 1 ---------
\begin{table}
\caption[ ]{Parameters of the adiabatic accretion disc models
(adiabatic index $\gamma_g=5/3$).
The radial distances $r_s$, $(r_s)_{crit}$, $r_{out}^{max}$
are given in units of the gravitational radius $r_g$.}
\begin{tabular}{lcccc}
\hline
 Model &~~~$\alpha$~~~&~~~$r_s$~~~&
~~~$(r_s)_{crit}$~~~&~~~$r_{out}^{max}$~~~\\
\hline
$A1$ & $0.1$ & $2.485$ & $2.485$ & $24.0$ \\
$A2$ & $   $ & $2.5  $ & $     $ & $16.8$ \\
$A3$ & $   $ & $2.55 $ & $     $ & $11.1$ \\
\hline
$B1$ & $0.01$ & $2.165$ & $2.165$ & $85.7$ \\
$B2$ & $    $ & $2.175$ & $     $ & $53.3$ \\
$B3$ & $    $ & $2.2  $ & $     $ & $29.0$ \\
$B4$ & $    $ & $2.25 $ & $     $ & $17.1$ \\
$B5$ & $    $ & $2.3  $ & $     $ & $13.0$ \\
\hline
$C1$ & $0.001$ & $2.09 $ & $2.09$ & $148.0$ \\
$C2$ & $     $ & $2.1  $ & $    $ & $ 73.0$ \\
$C3$ & $     $ & $2.125$ & $    $ & $ 37.4$ \\
$C4$ & $     $ & $2.15 $ & $    $ & $ 25.5$ \\
$C5$ & $     $ & $2.2  $ & $    $ & $ 15.7$ \\
\hline
$D1$ & $10^{-5}$ & $2.05$ & $2.05$ & $245.0$ \\
$D2$ & $       $ & $2.1 $ & $    $ & $ 35.6$ \\
$D3$ & $       $ & $2.15$ & $    $ & $ 18.6$ \\
\hline
\end{tabular}
\end{table}

\begin{figure}
 \hbox{\psfig{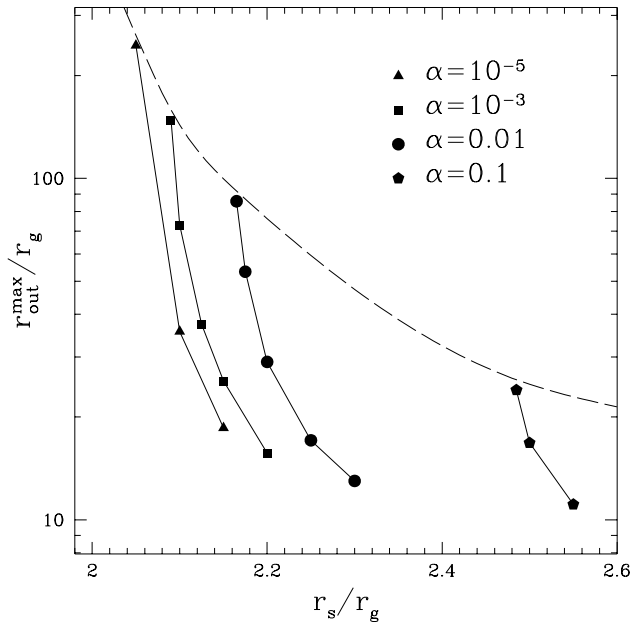}}
 \caption{
The dependence of maximal radial extension
$r_{out}^{max}$ of the adiabatic accretion
disc models (adiabatic index $\gamma_g=5/3$) on the position of the transonic
radius $r_s$.
The data are shown for
the viscosity parameter $\alpha=10^{-5}$, $10^{-3}$, $0.01$ and $0.1$
(triangles, squares, circles and pentagons, respectively).
In the region above the dashed line, slim disc solutions
do not exist.
}
\end{figure}

Table~1 summerizes the basic parameters of the computed models with
$\alpha=10^{-5}$, $10^{-3}$, $0.01$ and $0.1$.
In these models $r_{out}$ was set approximately equal to $r_{out}^{max}$.
Fig.~1 shows the dependence of $r_{out}^{max}$ on the position of the
sonic radius for the solutions listed in Table~1, with the dashed line 
showing roughly the boundary of the region in the
($r_s$, $r_{out}^{max}$) space where solutions of the first type exist.
More extended discs correspond to smaller $\alpha$ and smaller $r_s$.
Fig.~2 shows the disc shapes $H(r)$ for the largest discs (with
$r_s=(r_s)_{crit}$)
for the values of $\alpha$ listed above 
(models $A1$, $B1$, $C1$ and $D1$).
The shapes of these discs are very similar: the maximum vertical thickness
occurs at around $0.65 r_{out}^{max}$ and the outer edge is very sharp.
Figs~3, 4, and 5 show the variation with $r$ of the angular momentum
$\cal L$, the radial velocity $V$ and the equatorial pressure $p$,
for solutions with different values of $\alpha$ and $r_s$ (corresponding
to the models listed in Table~1).

\begin{figure}
 \hbox{\psfig{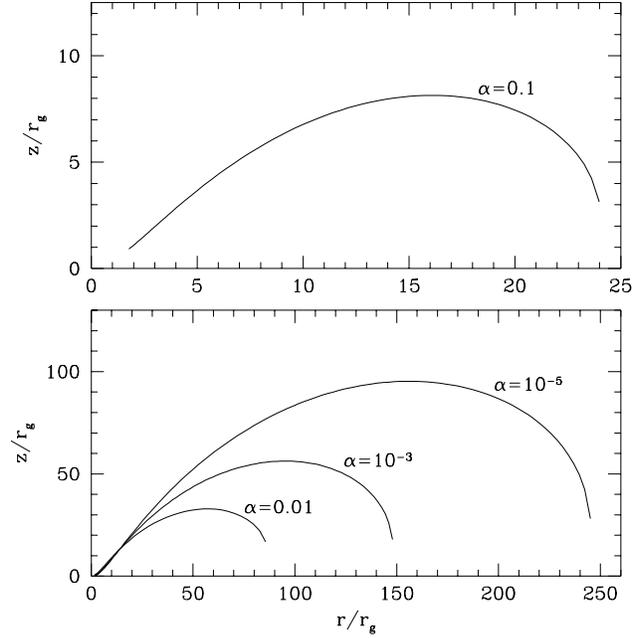}}
 \caption{
The shapes of the adiabatic accretion discs ($\gamma_g=5/3$)
for four values of the viscosity parameter
$\alpha=0.1$, $0.01$, $10^{-3}$ and $10^{-5}$.  The models presented have
the maximal radial extensions for the given $\alpha$, and correspond to
the models $A1$, $B1$, $C1$ and $D1$, listed in Table~1.
}
\end{figure}

At the outer edge, all of the models have low values of $V$ and $c_s$, and
the distribution of angular momentum has a super-Keplerian part. The
super-Keplerian rotation corresponds to a region where the pressure
increases with radius (see Fig.~5).  The pressure increases due to the
rapid decrease of the disc thickness $H$ at the outer boundary and the
corresponding increase in the matter density. The super-Keplerian rotation
compensates the gravitational force and the gradient in the gas pressure,
which are both directed inwards in this region.  The presence of
super-Keplerian rotation does not depend on the particular value taken for
${\Omega}_{out}$. We tried a variety of values, both Keplerian and
non-Keplerian, but always found the presence of super-Keplerian rotation.
Note, that the super-Keplerian rotation often occurs in a region where a
sharp transition in qualitative properties of the accretion flow takes
place:  the two sub-Keplerian flows on either side of the super-Keplerian
part are very different. For example, super-Keplerian rotation occurs at
the sharp, cusp-like inner edge of thick discs (Abramowicz, Calvani \&
Nobili 1980). In this case, the super-Keplerian part is sandwiched
between supersonic free-fall with negligible dissipation, and
centrifugally supported, viscously driven accretion flow. Super-Keplerian
rotation is also expected in the transition region between a standard
Shakura \& Sunyaev disc and an ADAF (Abramowicz, Igumenshchev \& Lasota
1997). We also note that the existence of super-Keplerian rotation at the
outer edge of some ADAF models can be seen in a few solutions previously
obtained by Honma (1996). 

\begin{figure*}
 \centerline{\hbox{\psfig{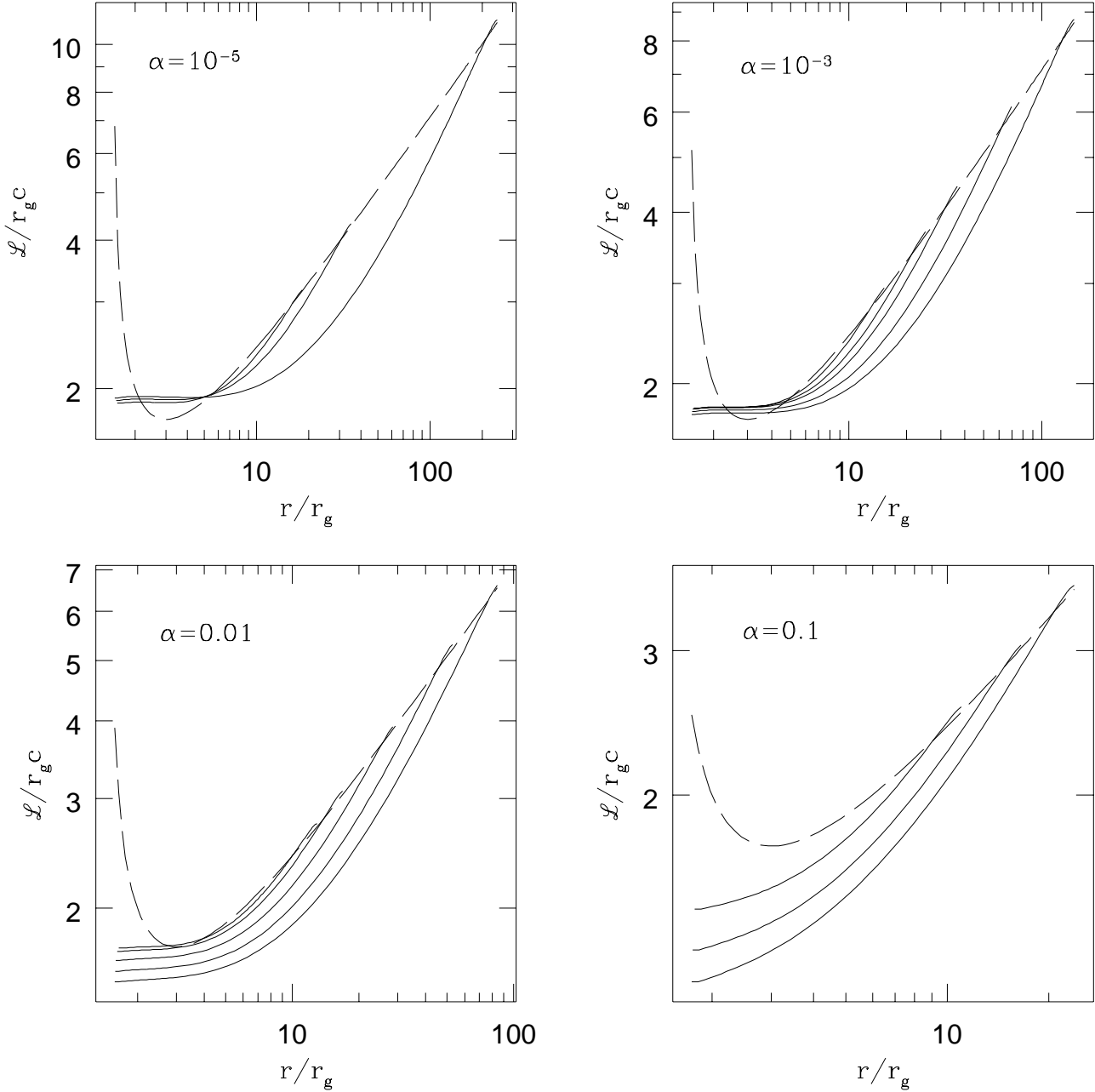}}}
 \caption{
Distributions of angular momentum in the adiabatic models
($\gamma_g=5/3$) for four
values of $\alpha$ (solid lines). The distribution of the Keplerian
angular momentum is shown for comparison (dashed lines).
For each
$\alpha$, the models have the values of the sonic radius $r_s$ and
the radial
extension $r_{out}=r_{out}^{max}$ given in Table~1.
}
\end{figure*}

\begin{figure*}
 \centerline{\hbox{\psfig{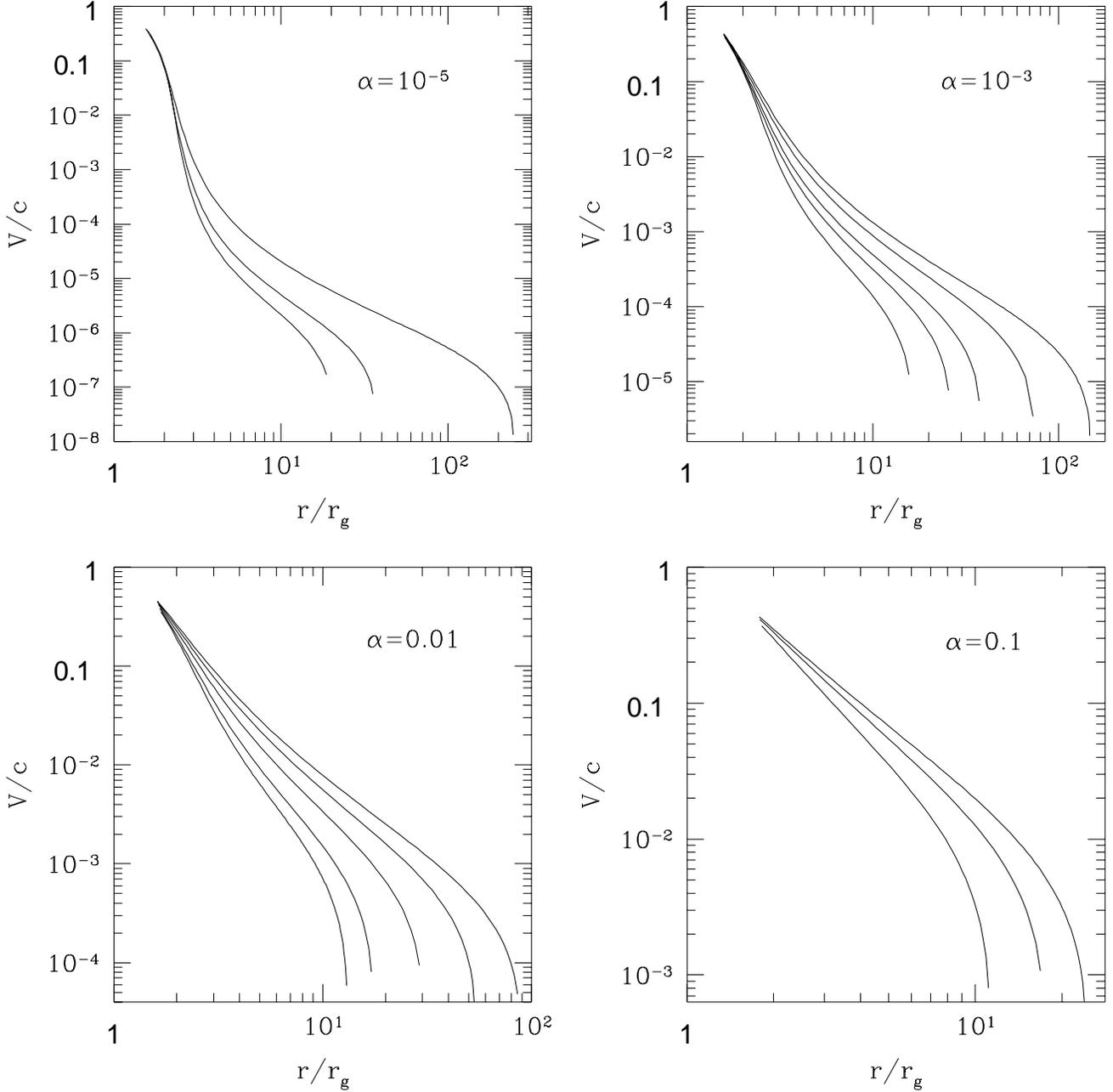}}}
 \caption{
Distributions of radial velocity in the adiabatic models with
$\gamma_g=5/3$ (see the caption to Fig.~3 for further explanation).
}
\end{figure*}

\begin{figure*}
 \centerline{\hbox{\psfig{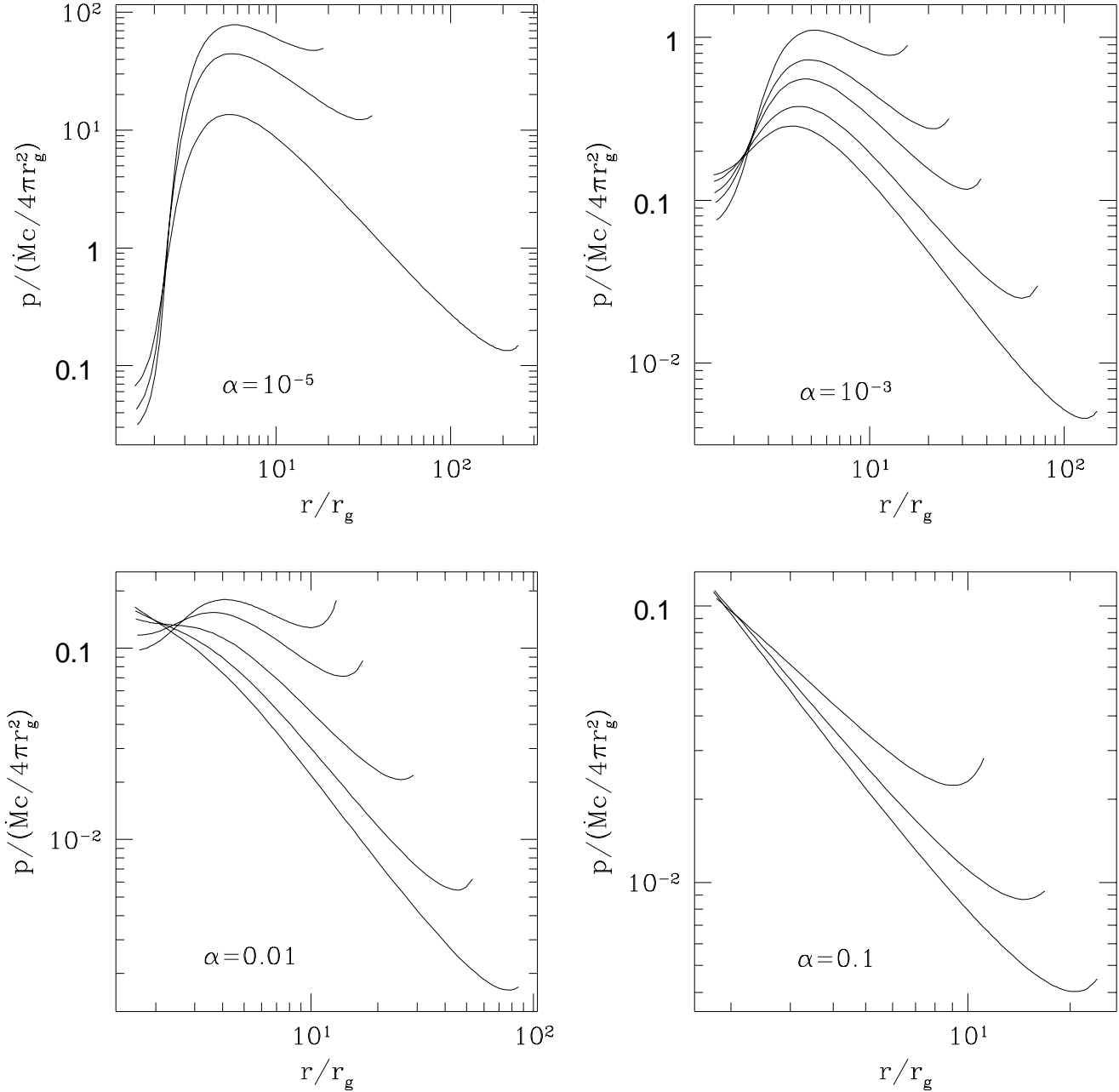}}}
 \caption{
Distributions of pressure in the adiabatic models with
$\gamma_g=5/3$ (see the caption to Fig.~3 for further explanation).
}
\end{figure*}

Just inside the outer boundary, the distribution of angular momentum
${\cal L}$ is almost a power law with $d\ln{\cal L}/d\ln r\approx 0.77$,
larger than the Keplerian value of $0.5$ (see Fig.~3). We note, that this
value is practically independent of $\alpha$ and depends only weakly on
$r_s$, if $r_{out}^{max}\ga 10 r_g$.  In this case, the rotational energy
of the accretion flow at the outer radii is being converted more
efficiently into internal energy than in the case of a standard thin disc. 
As a result, both the temperature and the thickness of the disc increase
rapidly inwards. A significantly sub-Keplerian rotation of this part of
the disc implies that the radial pressure gradient becomes important in
the balance of radial forces.

Near to the black hole, solutions with high (super-Keplerian) and low
(sub-Keplerian) angular momenta show qualitatively different
behaviour. This was first noticed long ago by Abramowicz
\& Zurek (1981).
%who explained that the high angular momentum solutions
%called the `disc-like') have sonic radius located at $r_s < r_{ms}$,
%where $r_{ms}$ is the radius of marginally stable orbit, while the low
%angular momentum solutions (called the `Bondi-like') have sonic
%radius located at $r_s > r_{ms}$.
When the boundary conditions are held
fixed, high and low values of the angular momentum near to the black hole
correspond to
solutions with small and large values of $\alpha$ respectively.
These two types of solution were referred to as `disc-like' and
`Bondi-like', respectively.
Differences between
the disc-like and Bondi-like solutions in the special case of
very geometrically-thin accretion flows have been discussed by
Muchotrzeb (1983), and later by several other authors, including
Matsumoto et al. (1984), ACLS88 and Narayan et al. (1997).
Our results show the existence of disc-like and Bondi-like
solutions in agreement with the general results of Abramowicz \&
Zurek (1981). In particular, we confirm a more recent specific example
of this general property, found by Narayan et al. (1997) in the
special case of ADAFs: for $\alpha\la 0.01$, solutions are of the
disc-like type, while for $\alpha\ga 0.01$ they are Bondi-like. For
the disc-like solutions with small $\alpha$, the sonic radius is
located near to the marginally-bound radius, $2r_g<r_s\la 2.3 r_g$, and
the distribution of angular momentum has a super-Keplerian part around the
marginally-stable radius (see Figs~1 and 3). For the Bondi-like
type, corresponding to large $\alpha$, the sonic radius is located closer
to $r_{ms}$, and the angular momentum is substantially sub-Keplerian
everywhere in the region.  Note, that our Bondi-like solutions have
$r_s<r_{ms}$.
We did not find solutions with $r_s> r_{ms}$ in case of
$\alpha\leq 0.1$.  The existence of solutions with $r_s\ga r_{ms}$ is
questionable because the maximum radius of the disc $r_{out}^{max}$ 
decreases quickly with increasing sonic radius for any $\alpha$
(see Fig.~1).

\noindent
Differences between these two classes of solution can also be seen in
the radial distribution of the pressure. Disc-like models have pressure
maxima outside the transonic point while these maxima disappear
in Bondi-like models (see Fig.~5).

Models with adiabatic index $\gamma_g=4/3$ only have properties which are
qualitatively similar to those with $\gamma_g=5/3$ if the viscosity is
high ($\alpha\ga 0.01$). In particular, all of our high viscosity models
with $H/r < 1$ and $\gamma_g=4/3$ are limited in size. It could be,
however, that there are also models which are unlimited in size but which
our numerical procedure was unable to find: for purely technical reasons,
it does not converge well for highly viscous models with large $r_{out}$. 

Models with $\alpha \la 0.001$ show a very different behaviour, however. 
Models with a particular value of the sonic radius, $r_s=(r_s)_0$ are not
limited in size and coincide asymptotically (for large radii) with the
corresponding self-similar solutions of Narayan \& Yi (1995). Models with
$r_s>(r_s)_0$ are limited in size with $r_{out}^{max}$ being a decreasing
function of $r_s$. For $r_s<(r_s)_0$ the solutions have $H/r>1$ in the
outer part and so do not satisfy the condition for the slim disc
approximation. 

\subsection{Bremsstrahlung models}

We now describe models with bremsstrahlung cooling, in particular, ones in
which the inner ADAFs are matched with outer SLE type solutions. In the
original SLE paper, ions and electrons were allowed to have different
temperatures, and the assumption of equality of the local electron heating
and cooling rates was used.  We use a one-temperature approximation for
plasma but, nevertheless, our bremsstrahlung solutions in the outer part
can be treated as being of SLE type in the sense that they are locally
cooled, i.e. $F^+\approx F^-$. 

\begin{figure*}
 \centerline{\hbox{\psfig{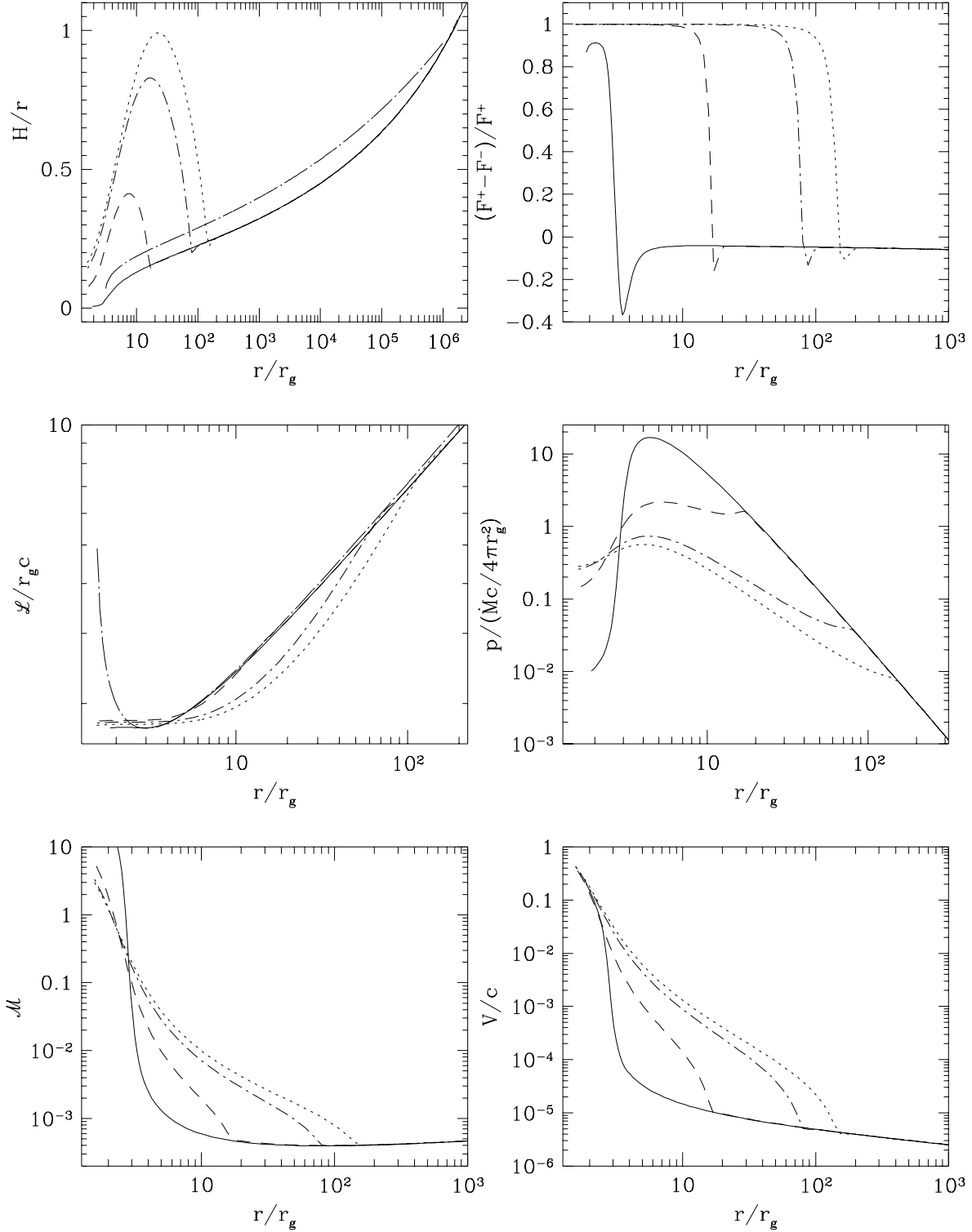}}}
 \caption{
Selected properties
of the bremsstrahlung models with $\gamma_g=5/3$, $\alpha=10^{-3}$,
$\dot{M}/\dot{M}_{Edd}=10^{-5}$ and $M=10\,M_\odot$.
The dotted,
dotted-short-dashed, short-dashed and solid lines represent the cases
$r_s/r_g=2.09$, $2.1$, $2.2$ and $2.7$, respectively.
The behaviour of the analytic SLE solution (3.1) is shown for
comparison by a dotted-long-dashed line in the aspect ratio $H/r$ plot.
A dotted-long-dashed line in the specific angular momentum ${\cal L}/r_g c$ 
plot shows the Keplerian angular momentum distribution.
The bottom left frame shows the run of the Mach number ${\cal M}$
as defined by equation (2.24).
The global solutions consist of two different parts with a relatively
narrow transition region.
At small $r$, the solutions behave like ADAFs, while
at large $r$ they behave like one-temperature SLE solutions.
The model with $r_s/r_g=2.7$ (solid line) has special properties:
it consists of the SLE solution everywhere except for the region inside
$3r_g$.
}
\end{figure*}

Fig.~6 shows the radial dependence of the relative thickness $H/r$, the
importance of the advective flux $(F^+-F^-)/F^+$, the angular momentum
$\cal L$, the equatorial pressure $p$, the Mach number $\cal M$, as
defined by equation (2.24), and the velocity $V$ for models with
$\gamma_g=5/3$, $\alpha=10^{-3}$ and $\dot{M}/ \dot{M}_{Edd} = 10^{-5}$.
Here $\dot{M}_{Edd} =4\pi GMm_p/c\sigma_T$ is the Eddington accretion
rate.  The black hole mass was taken to be $M=10 M_\odot$. In these models
we varied the position of the sonic radius. Dotted, dotted-short-dashed,
short-dashed and solid lines represent the cases $r_s/r_g=2.09$, $2.1$,
$2.2$ and $2.7$, respectively.  Inward of the transition radius, the
solutions become ADAFs with $(F^+-F^-)/F^+\simeq 1$, and they have all of
the properties of the adiabatic models discussed in the previous
paragraph.  The transition radius between the two types of solution
depends on $r_s$.  This dependence, as well as the limitations of the
minimum value of $r_s$, are very similar to those for $r_{out}^{max}$ (see
Fig.~1, square dots) and $(r_s)_{crit}$ in adiabatic models.  The model
with $r_s/r_g=2.7$ (solid lines in Fig.~6) has special properties, because
it becomes an ADAF just inside $3 r_g$, and the transition radius is very
close to the sonic radius. 

Outside the transition radius, each solution matches smoothly with
an SLE type solution, which is unique for given $\alpha$, $\dot{M}$ and
$M$, and does not depend on the position of the transonic point $r_s$ or
the location of the transition radius.  Note that we have not assumed
any extra physical effect that triggers the transition. The transition
occurs due to the standard accretion processes, described by the
standard equations for slim accretion discs.

\noindent
Of course, one my construct SLE solution analytically, taking
$\Omega=\Omega_K$, $F^+=F^-$ and zero viscous torque at $r=3r_g$,
%%%%%%%%%%%%%%%%%%%%%%%%%%%%%%%%%%%%%%%%%%%%%%%%%%%%%%%%%%%%%%%%%%%%%%%%%%%%%%%
\begin{equation}
  {H\over r}= 0.1\left({\dot{m}\over\alpha^2}\right)^{1/4}
   \left(1-\sqrt{{3r_g\over r}}\right)^{1/4} \left({r\over
   r_g}\right)^{1/8}, 
\end{equation}
%%%%%%%%%%%%%%%%%%%%%%%%%%%%%%%%%%%%%%%%%%%%%%%%%%%%%%%%%%%%%%%%%%%%%%%%%%%%%%%
where $\dot{m}=\dot{M}/\dot{M}_{Edd}$. The analytic SLE solution (3.1)
is shown in Fig.~6 by the dotted-long-dashed line. There is good
agreement between the analytic solution (3.1) and the numerical solution
outside the transition radius.  Both solutions have similar slopes
inside $\sim 10^5 r_g$. Fig.~6 shows that the SLE type solutions have
negative $(F^+-F^-)/F^+$ (see the discussion by Abramowicz 1996).
In the standard SLE solution $F^+=F^-$.

\begin{figure*}
 \centerline{\hbox{\psfig{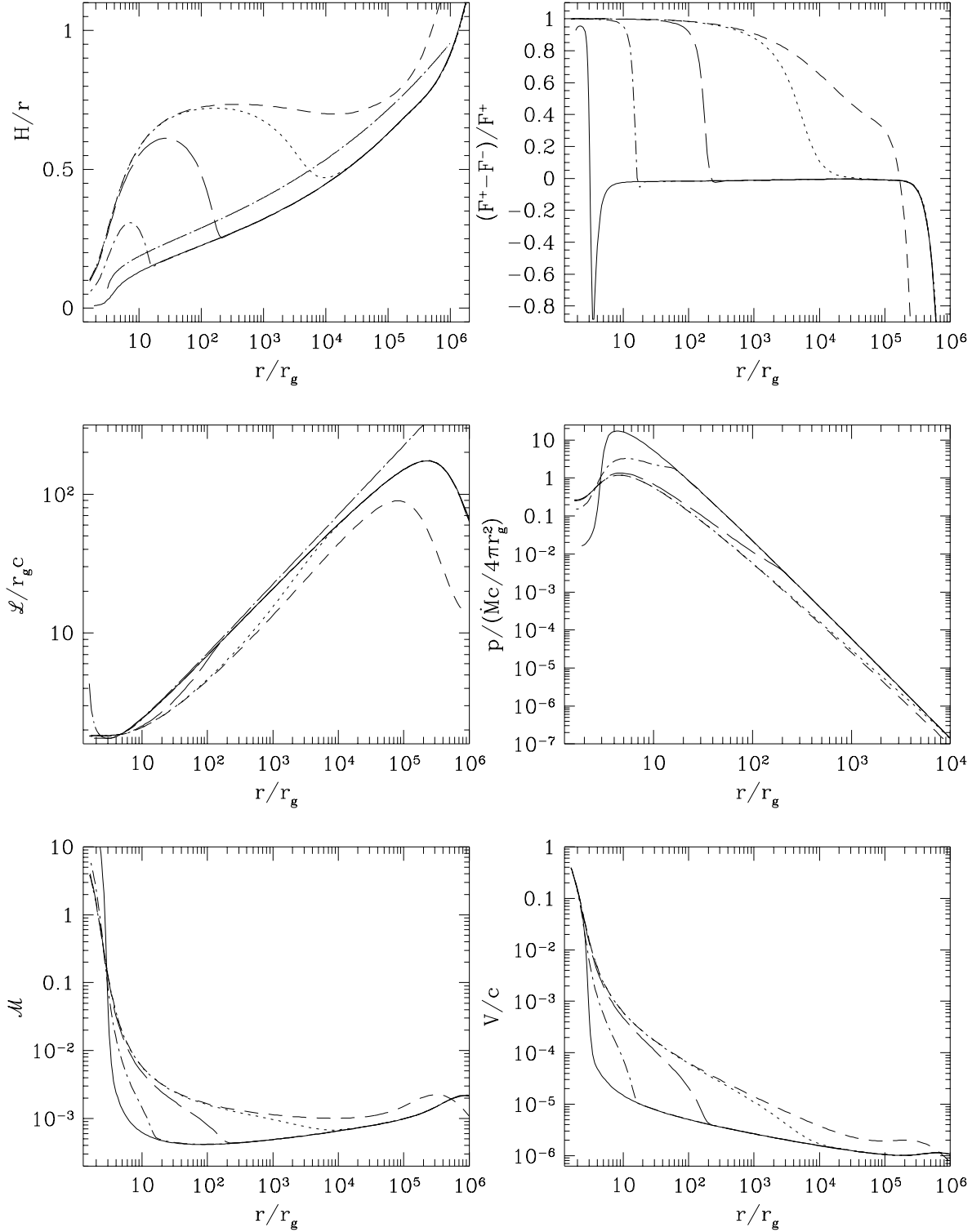}}}
 \caption{
Selected properties
of the bremsstrahlung models with $\gamma_g=4/3$, $\alpha=10^{-3}$,
$\dot{M}/\dot{M}_{Edd}=10^{-5}$ and $M=10\,M_\odot$.
Short-dashed, dotted, long-dashed, dotted-short-dashed and solid lines
represent the cases $r_s/r_g=2.13075$, $2.131$, $2.14$, $2.255$ and $2.7$,
respectively. See the caption to Fig.~6 for further explanation.
}
\end{figure*}

In Fig.~7 we show properties of the $\gamma_g=4/3$ models.  Short-dashed,
dotted, long-dashed, dotted-short-dashed and solid lines represent the
cases $r_s/r_g = 2.13075$, $2.131$, $2.14$, $2.255$ and $2.7$,
respectively. All of the models experience the ADAF-SLE transition. Note,
that in models corresponding to the short-dashed and dotted lines, the
ADAF part is very extended.  In contrast to the previous ($\gamma_g=5/3$)
case, we found an unphysical behaviour of the solutions for $r > 10^5
r_g$. In this radial range, the solutions have the angular momentum $\cal
L$ decreasing outwards, which is dynamically unstable. They also deviate
strongly from the thermal structure characteristic of the SLE solution,
having the cooling rate $F^-$ significantly larger than the heating rate
$F^+$. This behaviour does not depend either on the choice of the location
of the outer numerical boundary $r_{out}$, or on the value taken for
$\Omega_{out}$. 

\section{DISCUSSION AND CONCLUSIONS}

We have constructed numerical solutions for two classes of models of
optically thin accretion discs around black holes: adiabatic models, with
no radiative cooling, and models with bremsstrahlung cooling. 

All of the adiabatic models are of the ADAF type. Those with
$\gamma_g=5/3$ and $H/r\la 1$ everywhere are limited in size: their
radial extensions are smaller than about $10^2 r_g$. Independently of
whether viscosity is high or low, the vertical thickness $H$ of these
models tends to zero at the outer limiting radius. Models with
$\gamma_g=4/3$  and $H/r\la 1$ extend to radial infinity {\it only}
when the viscosity is low  (we constructed models with $\alpha\la
10^{-3}$) and when the location of the sonic point has exactly
a particular value $r_s=(r_s)_0$. Obviously, this is the eigenvalue
of the problem. It is therefore not surprising that these models have,
at large distances, the same properties as the models corresponding to
the self-similar solution of Narayan \& Yi (1995). For $r_s>(r_s)_0$
our models are radially limited. For discs with high viscosity
($\alpha\ga 10^{-2}$) we could not find 
disc-like solutions with unlimited radial size having $H/r\la 1$
everywhere.
We cannot conclude
that such solutions do not exist, due to numerical problems which our
code experiences in this case. Radially unlimited solutions that we
found in this range of parameters, all have the maximum value of
$H/r$ being an increasing function of the location taken for the outer
numerical boundary $r_{out}$. In particular, for large $r_{out}$ one
always has $H/r\gg 1$. This indicates that the one-dimensional approach
based on vertical integration which assumes $H/r\la 1$ is not
self-consistent and slim models for this type of disc cannot be
trusted.

Bremsstrahlung models consist of two parts which are joined by a
smooth transition region.
In some cases the transition region is relatively narrow
(models with $\gamma_g=5/3$), but in other cases it can be larger than
the ADAF.
Inside the transition radius, the
solution is very similar to the ADAF-type adiabatic solution. Outside it,
the solution is of the SLE type, with the entropy of the matter increasing
with radius. Our numerical SLE-type solutions coincide closely with the
analytical ones.  Obviously, though, this kind of global solution cannot
exist in nature, because the outer SLE-type disc is thermally unstable.

The most interesting property of the ADAF solutions found in this paper is
their behaviour near to the transition radius. It was known previously
(see e.g. discussion given by Chen et al. 1995) that the
ADAF-type accretion flows cannot extend to arbitrarily large radii~---
they may extend out to $\sim 10^6 r_g$.  The limitation is due to the
fact that at large radii the accretion time scale exceeds the free-free
cooling time. This case is represented by the model shown by the 
short-dashed line in Fig.~7. At the range of radii from $10^2 r_g$ to around
$10^4 r_g$ the short-dashed model is an ADAF, which is similar to
the self-similar solution of Narayan \& Yi (1995), where $H/R\approx
const$ and $\Omega/\Omega_K\approx const$. At $r\ga 10^5 r_g$ the
radiative cooling becomes important in the energy balance and the
solution ceases to be of the ADAF type. However, the end of the
ADAF-type part for the other models shown in Figs~6 and 7 
must be of a different nature. It cannot be connected to the
cooling mechanisms, because both adiabatic and non-adiabatic models
show very similar behaviour. One should note that the position of the
end of the ADAF is determined by the value of the parameter $r_s$: a
larger $r_s$ corresponds a smaller transition radius.

The absence of the radially-unlimited ADAF solutions for
$\gamma_g=5/3$ could be connected with the fact that there are no
disc-like asymptotic self-similar solutions for $\gamma_g=5/3$.  The
case $\gamma_g=5/3$ is singular (see Narayan \& Yi 1995). On the other
hand, we note,
that the analogous self-similar solutions derived from
the slim disc equations in the form given by Chen et al. (1997)
are free from the $\gamma_g =5/3$ singularity. Our slim disc equations
are similar to the ones used by Narayan \& Yi (1995), and therefore it
is natural that our results agree with theirs. The question, however,
remains of why different versions of the slim disc equations produce
qualitatively different solutions at large radial distances $r\ga 10^2
r_g$? Are seemingly `formal' details of the vertical integration
crucial for the physics of slim disks? Is the slim disc approach
questionable in application to ADAFs? The best way to answer these
questions would be by constructing fully two dimensional,
non-stationary models of ADAFs.

There is a less important, but related, problem here.  We have not
obtained global solutions with shocks like those reported by Chakrabarti
and
collaborators (see Chakrabarti 1996) despite the very similar physics
and boundary conditions assumed. Some of our solutions have transition
regions which are narrow, but smooth with their inner structure being 
numerically well resolved. These narrow transitions are definitely {\it
not}
shocks. It is likely that this will be taken as an argument in the
recent discussion `shocks or no shocks' but we do not want to join
in with this ourselves because we are convinced that the
discussion is artificially inflated and of only academic interest.
It does not touch a really interesting point here that one should be
aware of. Several authors have recently studied
general properties of transition regions between ADAFs and thin discs
(see e.g. Regev, Lasota \& Abramowicz 1997; Abramowicz et
al. 1997b) and concluded that the physical conditions in the
transition region are very complex, and cannot be described by any
simple one-dimensional, stationary approach.
%We have not found solutions that at smaller radii are of the ADAF type
%and at some intermediate radius experience a transition to the
%standard (stable) thin accretion disc (Shakura, 1972). This may be taken
%as an indication that for such transitions additional physical processes,
%for example evaporation, is indeed necessary.
Details of the vertical
structure and dissipative processes are fundamental for the
structure of the transition
region and must be fully taken into account. Thus, one
should first construct a more {\it physical} model of the transition
region, before speculating about its nature.

\section*{Acknowledgments}

We thank Kees Dullemond, Jean-Pierre Lasota,
Roberto Turolla and Craig Wheeler for
stimulating discussions.  
We thank John Miller for his helpful comments on the paper.
This work was partially supported by the
Swedish Natural Science Research Council, by
Nordita's Nordic Project {\it Non-linear phenomena in accretion discs
around black holes}, by the Danish Natural Science Research Council
through grant 11-9640-1, and by the Danish National Research Foundation
through its establishment of the Theoretical Astrophysics Center.

\bsp

\label{lastpage}

\end{document}